\documentclass[%
 reprint,
 amsmath,amssymb,
 aps,
 pra,
]{revtex4-2}

\usepackage{lmodern}
\usepackage{graphicx}% Include figure files
\usepackage{dcolumn}% Align table columns on decimal point
\usepackage{bm}% bold math
\usepackage{hyperref}% add hypertext capabilities
\hypersetup{colorlinks=true, citecolor=blue, urlcolor=blue, linkcolor=blue}
\graphicspath{{Figures/}}
\usepackage{blindtext}

\newcommand{\nabp}{\nabla_{\perp}}
\newcommand{\Hp}{\mathbf{H}^{+}}

\newcommand{\Dr}{D_{\mathbf{r}}}

\begin{document}

\preprint{APS/123-QED}

\title{Rotating and spiralling spatial dissipative solitons of light and cold atoms}% Force line breaks with \\

\author{Giuseppe Baio}
\email{giuseppe.baio@strath.ac.uk}
\author{Thorsten Ackemann}
\author{Gian-Luca Oppo}
\author{Gordon R. M. Robb}
\author{Alison M. Yao \vspace*{.1cm}}

\affiliation{
 SUPA and Department of Physics, University of Strathclyde, Glasgow G4 0NG, Scotland, United Kingdom
}%

\date{\today}% It is always \today, today,

\begin{abstract}
% Clouds of cold neutral atoms driven by a coherent beam of light exhibit self-structured states, transversally with respect to the beam axis. For parameter regimes where atomic density modulations induced by optomechanical forces are prominent, dynamical structures of light and atoms with hexagonal ordering are observed, where atomic transport can be engineered by means of phase structured input fields. As a consequence of their subcritical character, localized cavity solitons can be excited below threshold. Moreover, it is shown that by appropriately tailoring the phase structure, one can induce complex motion of optomechanical cavity solitons such as rotating or spiralling trajectories. We also discuss the feature of soliton-soliton interactions with the cold-atom optomechanical nonlinearity and the stability of multiple bound states and interacting soliton chains.

Clouds of cold neutral atoms driven by a coherent light beam in a ring cavity exhibit self-structured states transversely with respect to the beam axis due to optomechanical forces and the back action of the atomic structures on the beam. Below the instability threshold for extended hexagonal structures, localized soliton-like excitations can be stable. These constitute peaks or holes of atom density, depending on the linear susceptibility of the cloud. Complex rotating and spiralling motion of coupled atom-light solitons, and hence atomic transport, can be achieved via phase gradients in the input field profile. We also discuss the stability of rotating soliton chains in view of soliton-soliton interactions. The investigations are performed in a cavity scheme but expected to apply to other longitudinally pumped schemes with diffractive coupling.

\end{abstract}

\maketitle

%\tableofcontents

\section{\label{sec:level1}Introduction}

Dissipative solitons are stable localized excitations of nonlinear systems that include drive and dissipation, generalizing the concept of solitary wave-like solutions (solitons) to the case of non-integrable systems \cite{akhmediev2008dissipative}. Transverse optical systems involving diffractive feedback or optical cavities provide suitable platforms for observing spatial dissipative solitons within the nonlinear optics domain \cite{ackemann2009fundamentals}. A feature that is well described by mean-field models is that optical dissipative solitons can be switched on and off by means of tightly focused addressing pulses \cite{purwins2010dissipative}. Therefore, multiple spatial peaks of the optical field can be excited locally in the system \cite{mcsloy2002computationally, vladimirov2002two}, exhibiting characteristic homoclinic snaking branches \cite{burke2007snakes}.

Schemes involving cold atomic gases in longitudinally pumped cavities or single-mirror systems have been recently investigated theoretically and experimentally \cite{ackemann2021self}, and shown to achieve light-atom self-structuring, where the relevant coupling involves optomechanical (dipole) forces \cite{Tesio2012, Labeyrie2014a, Tesio2014a, Robb2015}, electronic \cite{camara2015optical, Labeyrie2016} and magnetic transitions \cite{kresicCP, labeyrie2018magnetic, krevsic2019inversion, ackemann2021coupling}. In particular, the collective nature of optomechanical self-structuring has provided insight into several aspects of cold and ultracold atom physics such as crystallization \cite{Ritsch2013, Ostermann2016, gopalakrishnan2009emergent}, supersolidity with continuous symmetry breaking \cite{gopalakrishnan2010atom, mottl2012roton, leonard2017supersolid}, photon-mediated interactions with tunable range \cite{vaidya2018tunable}, and structural transitions between crystalline configurations \cite{li2020measuring, baio2021multiple}.

Transverse optical pattern forming dynamics involving orbital angular momentum (OAM) in the input beam was analyzed for a photorefractive medium \cite{caullet2011pattern, caullet2012vortex}, domain walls in optical parametric oscillators \cite{oppo2001characterization}, and dissipative solitons in semiconductor microcavities \cite{kheradmand2003rotating}. However, a systematic treatment of such phenomena  was developed only recently in Ref.~\cite{yao2019control} for a Kerr cavity, including polarization structuring effects. 

In this work, we focus on a cold atom optomechanical cavity scheme as introduced in Ref.~\cite{Baio2020}. There, it was shown that the rotational dynamics of self-organized light-atom ring-lattices is capable of sustaining robust atomic transport, induced by the input beam carrying OAM. Cavity solitons (CSs) of light and atom density were theoretically predicted with a purely optomechanical nonlinearity in Ref.~\cite{Tesio2013}. Here we show that a phase structured input can be used to engineer transport of self trapped atoms along complex trajectories, such as rotating or spiralling motion in the transverse plane, and to probe multi-soliton interactions resulting in stable soliton clusters. These considerations are also expected to apply to the dark and bright solitons in cold atoms predicted in a single-feedback-mirror (SFM) scheme \cite{baio2021multiple}. 

The paper is organized as follows. In Sec.~\ref{sec:level2} we review the features of mutually self-focused light-density soliton in contrast to optomechanical CSs. In Sec.~\ref{sec:level3} we discuss the motional dynamics of CSs induced by the phase gradients and in Sec.~\ref{sec:level4} the existence of different soliton configurations depending on the cloud susceptibility. Finally, in Sec.~\ref{sec:level5}, we address the properties of CS interactions in rotating chains. 

\section{\label{sec:level2}The model}

\begin{figure}
\hspace{-.45cm}\includegraphics[scale=1.3]{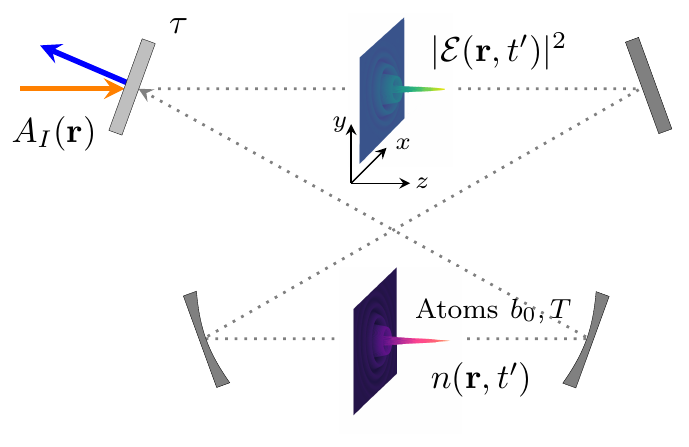}
\caption{Optomechanical single-mode ring cavity in a bow-tie configuration. The intra-cavity field $\mathcal{E}(\mathbf{r},t')$ is driven by a phase structured beam of amplitude $\mathcal{A}_{\mathrm{in}}(\mathbf{r})$, where $\mathbf{r} = (x,y)$ is the transverse coordinate. We assume one imperfect mirror with transmittivity $\tau$. An ensemble of overdamped laser-cooled two-level atoms of temperature $T$ and optical density at resonance $b_0$ is placed within the cavity, coupled to the optical field via a Smoluchowski equation for the atom density $n(\mathbf{r},t')$. CSs arise from the bi-stability of patterned and homogeneous states below threshold \cite{ackemann2009fundamentals, Tesio2013}. }
\label{cavitysol}
\end{figure}

One of the mechanisms allowing the formation of spatial solitons in light-atom systems relies on a collective self-focusing effect due to modulations of the atom density. The concept of mutual self-focusing occurring in  coupled light and atomic beams was introduced first by Klimontovich and Luzgin in Ref.~\cite{klimontovich1979possibility}.  The total index of refraction for a two-level atomic medium is obtained from its nonlinear susceptibility as follows \cite{Labeyrie2014a}:
\begin{align}
\nu(\mathbf{r}, s) =  1 - \frac{3 \lambda^3}{8 \pi^2}\frac{\Delta}{(1+\Delta^2)}\frac{n(\mathbf{r}, s)}{1+s(\mathbf{r})},
\end{align}
where $n(\mathbf{r}, s)$ is the spatially modulated atom density, $s(\mathbf{r})$ is the atomic saturation parameter, $\Delta = 2\delta/\Gamma$ the light-atom detuning in units of half-linewidth and $\lambda$ the light wavelength. By assuming a Gibbs equilibrium state for $n(\mathbf{r}, s)$ on the integration domain $\Omega$ as follows:
\begin{align}
    n_{\textrm{eq}}(\mathbf{r}) = \frac{\exp{\left[-\sigma s(\mathbf{r}) \right]}}{\int_{\Omega}\exp{\left[-\sigma s(\mathbf{r}) 
    \right]}d\mathbf{r}},
    \label{canonical}
\end{align}
where $\sigma= \hbar\delta/2k_{B}T = \hbar\Delta\Gamma/4k_{B}T$ represents an optomechanical coupling constant, one easily derives the Klimontovich-Luzgin condition for mutual self-focusing with red atomic detuning $\Delta <0$ only, namely \cite{klimontovich1979possibility}: 
\begin{align}
	\frac{\exp{[-\sigma s(\mathbf{r})]}}{1+s(\mathbf{r})} >1 \quad\textrm{for}\quad \sigma < 1.
\end{align}
Wang and Saffman showed that the same condition allows for stable spatial soliton solutions in the paraxial propagation of an optical field through an atomic cloud \cite{Saffbook, wangthesis}. Assuming instead a ring cavity configuration, such as the one sketched in Fig.~\ref{cavitysol}, leads to the following nonlinear coupled  model for the slowly varying envelope of the optical field $\mathcal{E}$ in the mean field and low saturation approximations \cite{lugiato1988stationary}:
\begin{align}
    \partial_{t'} \mathcal{E} =  -(1+i\theta)\mathcal{E} + \mathcal{A}_{\mathrm{in}}(\mathbf{r})  -2 i \mathcal{C} \Delta\, n\,\mathcal{E}  + i\nabp^{2}\mathcal{E}, \label{ringcaveq} 
\end{align}
where $t' = \kappa t$ is an adimensional time variable (in units of cavity lifetime $\kappa$), $\theta$ is the detuning between the pump and the closest cavity resonance, $\mathcal{A}_{I}(\mathbf{r})$ a spatially dependent pump rate, and $\mathcal{C} = b_0/2\tau(1+\Delta^2)$ the cavity cooperativity parameter, describing the susceptibility strength at fixed $\Delta$. The atom density $n(\mathbf{r},t')$ obeys the Smoluchowski equation in the strong friction limit \cite{Tesio2012, Tesio2013}: 
\begin{align}
    \partial_{t'} n = \sigma D_{\mathbf{r}}\nabp\cdot\left[n\,\nabp |\mathcal{E}|^2\right]+D_{\mathbf{r}}\nabp^{2}n,
    \label{smoluch2}
\end{align}
where $D_{\mathbf{r}}$ is the spatial diffusion constant \footnote{Note that this can be controlled externally by means of optical molasses beams via the momentum damping rate, assumed such that the overdamped condition is satisfied.}. Optomechanical transport is generated by the dipole potential, indicated by the gradient terms in Eq.~($\ref{smoluch2}$). 
Similarly to the case of mutual filamentation instabilities of light and matter beams in Ref.~\cite{saffman1998self}, the coupled system of Eqs.~(\ref{ringcaveq}) and (\ref{smoluch2}) is shown to have a modulation instability leading to optomechanical structure formation for a plane wave pump, namely $\mathcal{A}_{\mathrm{in}}(\mathbf{r}) = A_{\mathrm{in}} $, when the following minimum threshold condition is satisfied \cite{Tesio2012, Tesio2013}:
\begin{align}
I = |A_{\mathrm{in}}|^2 \geq I_0 =  \frac{1}{2\mathcal{C}\Delta\sigma} .  
\end{align}

The optomechanical instability typically results in the formation of a positive hexagonal phase $\mathbf{H}^{+}$ of the cavity field $\mathcal{E}(\mathbf{r},t)$ together with correlated (anticorrelated) density $n(\mathbf{r},t')$ in a $\mathbf{H}^{+(-)}$ for red (blue) detuning $\Delta < 0 $ ($\Delta > 0 $) \cite{Tesio2012, Baio2020}. Deviations from the effective-Kerr approximation and structural transitions are expected to arise in the strong detuned regime (at fixed $b_0$), namely, when the cloud susceptibility $\mathcal{C}\Delta$ is lower than a critical value  \cite{baio2021multiple}. Finally, as shown in Ref.~\cite{Tesio2013}, the subcritical bi-stability of hexagonal phases allows for coupled light-density dissipative solitons beyond the mutual self-focusing condition. In the rest of the paper, we focus on parameter regions where the atom density exhibits a $\Hp$ phase, since the localized structures correspond to peaks of self-trapped atoms. As shown later in Sec.~\ref{sec:level4}, this is also possible for blue-detuned atoms, where self-trapping and cooling conditions in a cavity QED system have been explored recently in Ref.~\cite{jungkind2019optomechanical}.

\section{\label{sec:level3} Soliton dynamics with structured phase profiles}

\begin{figure}
\hspace*{-.2cm}\includegraphics[scale=.21]{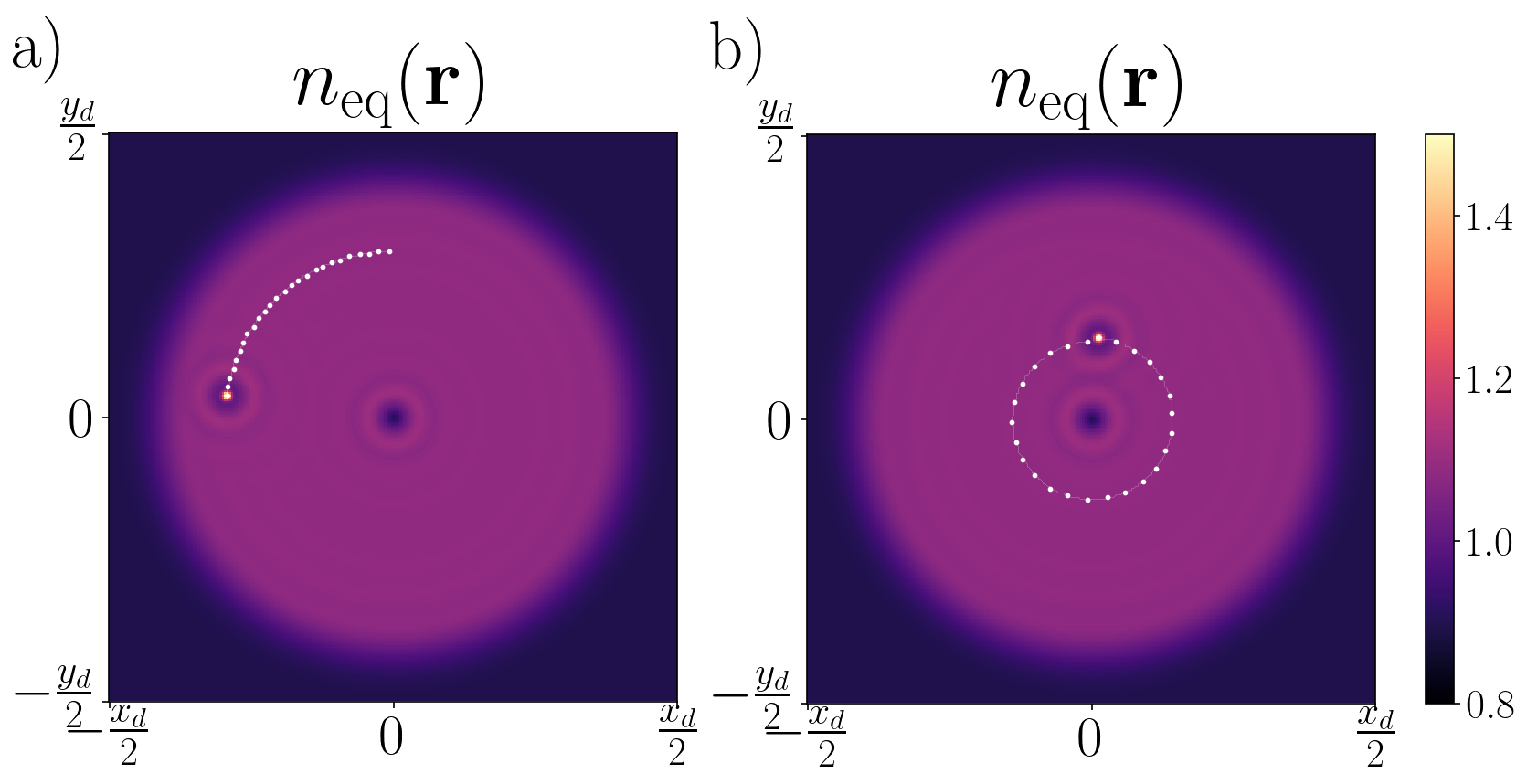}
\caption{Rotation of bright density optomechanical CSs for $\Delta < 0$, $\kappa t_{\mathrm{max}} = 300$, with OAM index $l=1$ at different radii (The rotation is counterclockwise for $l>0$.). The white dots track the peak position at different times.~(a) A localized peak of both the cavity field and atom density, initially defined at a position $\mathbf{r}_0 = (0,y_0)$ with $y_0 >y_d/4$, covers approximately a quarter of its orbit. (b) For an inner radial initial position $y_0<y_d/4$, the CS achieves a faster rotation speed (see Eq.~(\ref{driftvel})), completing a cycle within $\kappa t_{\mathrm{max}}$. Model parameters chosen as follows: $I/I_0 \approx 0.668$, $\theta = 5.1$,  $\mathcal{C} \Delta = - 2.25$, $\sigma = 25$.}
\label{rot2}
\end{figure}
Rotational or spiralling motion of localized structures induced by an azimuthal phase twist was predicted for solitons supported by Bessel lattices in cubic media in Ref.~\cite{kartashov2004rotary, kartashov2005soliton}, and observed experimentally in Ref.~\cite{wang2006observation}. We start here by considering the simplest scalar phase structured input profile carrying OAM, namely:  
\begin{align}
    \mathcal{A}_{\mathrm{in}}(\mathbf{r}) = A_{\mathrm{in}}(r)\exp(il\phi),
\end{align}
where the amplitude $A_{\mathrm{in}}(r)$ is a radial function and $\mathbf{r} = (r,\phi)$ represent the transverse position expressed in polar coordinates. $A_{\mathrm{in}}(r)$ is assumed as the following hyperbolic tangent "tophat":
\begin{equation}
A_{\textrm{in}}(r)= \frac{\sqrt{I}}{2}\left\{1-\textrm{tanh}[\xi(r-\rho_{0})]\right\},
\label{tophat}
\end{equation}
with controllable steepness $\xi$ and size $r_{0}$ \cite{eslami2014complex}. Note that no radial modulation is present, in contrast with Ref.~\cite{kartashov2004rotary, kartashov2005soliton}. As for the well known cases of Laguerre-Gaussian or Bessel beams, the purely azimuthal phase factor $\exp(il\phi)$, with $l\in\mathbb{Z}$, generates a nontrivial vortex structure with phase singularity at $\mathbf{r} = 0$ \cite{yao2011orbital}. Finally, the CS is seeded as a localized pattern peak defined on top of Eq.~(\ref{tophat}).    

Numerical simulations of the 2D optomechanical CS dynamics, in the model described by Eqs.~(\ref{ringcaveq}), together with the Gibbs distribution for the atom density in Eq.~(\ref{canonical}), are performed by means of a split-step method, with a spatial domain size of 10 critical wavelengths \footnote{The critical wavelength $\Lambda_c = 2\pi/q_c$ for the optomechanical ring cavity model can be found in Ref.~\cite{Baio2020}.} in a $256^2$ grid and time-step $\delta t' = 10^{-3}$. The purely azimuthal rotation case is shown in Fig.~\ref{rot2} for red detuning $\Delta < 0$, where the bright self-trapped density peak is observed drifting along perfectly circular trajectories with different rotation speeds at different radii. As for the case of optical bullet holes in the purely absorptive model, the drift velocity is determined by the input phase gradient \cite{Firth1996a} and, thus, the observed rotation speed scales with the transverse radius $r$ as follows \cite{yao2019control, Baio2020}:
\begin{align}
    \mathbf{v}_{\textrm{dr}}(\mathbf{r}) \propto 2l\nabp\phi  = \frac{2l}{r} \hat{\phi},
    \label{driftvel}
\end{align}
as visible from the tracked position of the maximum of the atom density peak $n_{\textrm{eq}}(\mathbf{r},t')$ in the transverse domain, measured at regular intervals of $\kappa t = 10$. Note that, for the cold atom case, the exact proportionality is determined by the atomic diffusion timescales, which are shown to drag the rotation speed when $\Dr \rightarrow 0$ \cite{Baio2020}.

\begin{figure}
\hspace*{-.4cm}\includegraphics[scale=.21]{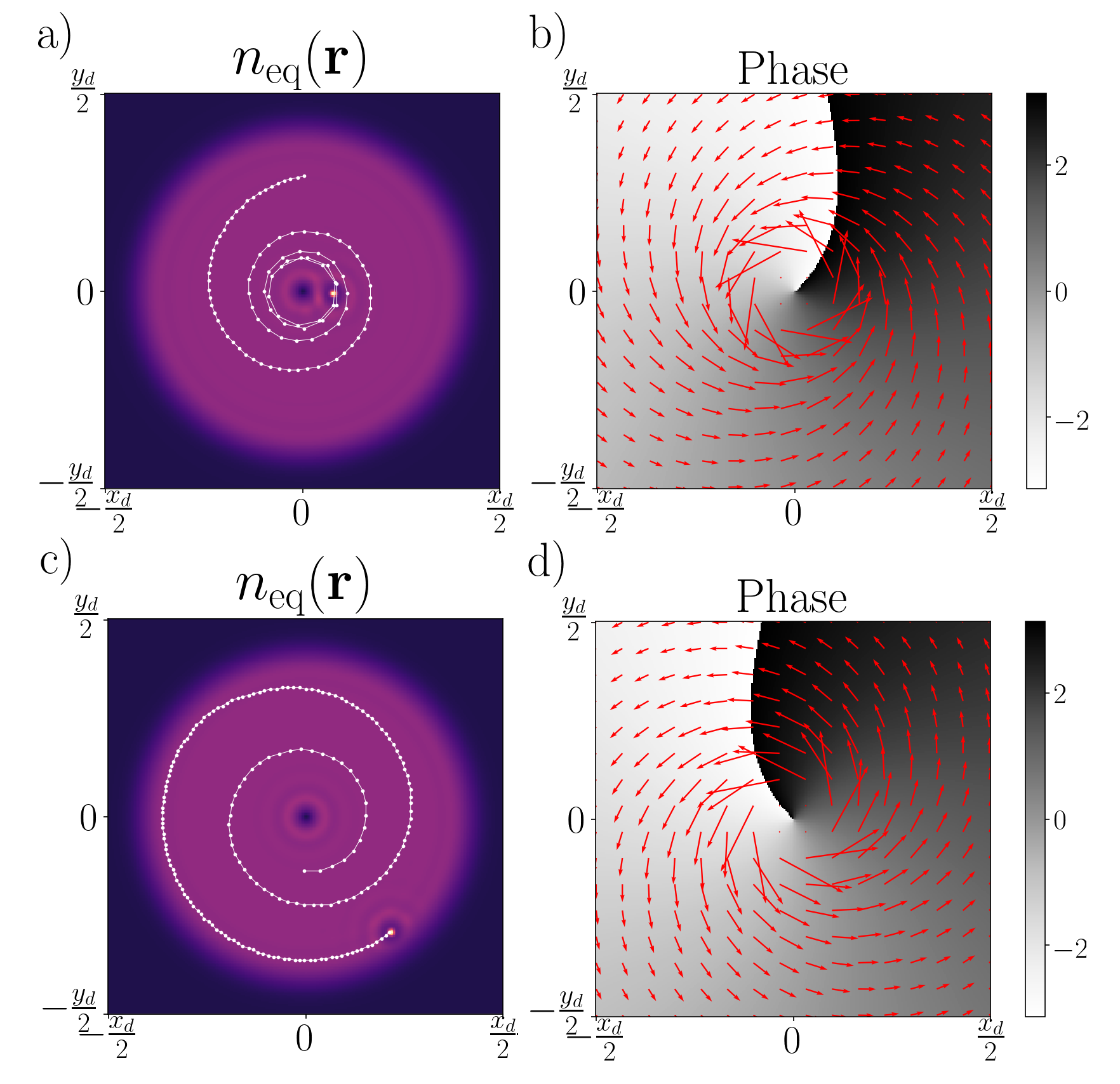}
\caption{Spiralling trajectories of optomechanical CSs for $\alpha \neq 0$ and  $\kappa t_{\mathrm{max}} = 750$. (a) Atom-density peak undergoing inward spiralling motion for  $\alpha  = 1.2$. (b) Total input phase $\alpha\psi(r) + \phi$ (OAM index $l=1$), including  gradient field, representing the soliton drift velocity. (c) Outward spiralling trajectory for $\alpha = -1.2$. (d) Corresponding phase and velocity field. Model parameters: $I/I_0 \approx 0.672$, $\theta = 5.1$,  $\mathcal{C} \Delta = - 2.25$, $\sigma = 25$.}
\label{spir2}
\end{figure}
Simultaneous control of angular and radial motion of the optomechanical CS can be achieved by means of an additional phase factor $\exp{\left[i\alpha\psi(r)\right]}$, where $\psi(r)$ is a concave function such that, e.g., when $\alpha > 0$, CSs are guided towards the center of the transverse domain, as the phase gradient field points towards the point $\mathbf{r}=0$. For convenience, $\psi(r)$ is chosen as in Eq.~(\ref{tophat}) with the same $\rho_0$ and a slower steepness $\xi$. The drift velocity in this case reads:
\begin{align}
    \mathbf{v}_{\textrm{dr}}(\mathbf{r}) = \alpha \partial_r \psi(r)\hat{r} +  \frac{2l}{r} \hat{\phi}.
    \label{driftspir}
\end{align}
The presence of radial correction for $\alpha \neq 0$ is shown here in Fig.~\ref{spir2}. The choice $\alpha > 0$ induces inward spiralling motion of the CS, until effective repulsive interactions close to the singular point $\mathbf{r} = 0$ take place, forcing the CS into a stable circular orbit, as shown by tracked evolution of the density peak in Fig.~\ref{spir2} (a). 
The inwards spiralling structure of the drift velocity  $\mathbf{v}_{\textrm{dr}}(\mathbf{r})$ is represented graphically in Fig.~\ref{spir2} (b), where the input phase and its gradient field are plotted together. The case $\alpha<0$ is shown in Fig.~\ref{spir2} (b)-(c), where the outward spiralling soliton is interacting this time with the outer boundary of the tophat, as shown by a slight change of direction in Fig.~\ref{spir2} (c). Eventually, the soliton is trapped in a circular trajectory at the maximum radius $\rho_0$ allowed by the tophat input profile \footnote{Note that the curvature of the spiral can be controlled by the concavity of $\psi(r)$ via the parameter $\alpha$.}. 

\section{\label{sec:level4} Dark light - bright atom density solitons}

%\subsubsection{Structural transitions among multiple phases}

\begin{figure}
    \centering
    \hspace*{-.7cm}\includegraphics[scale=.225]{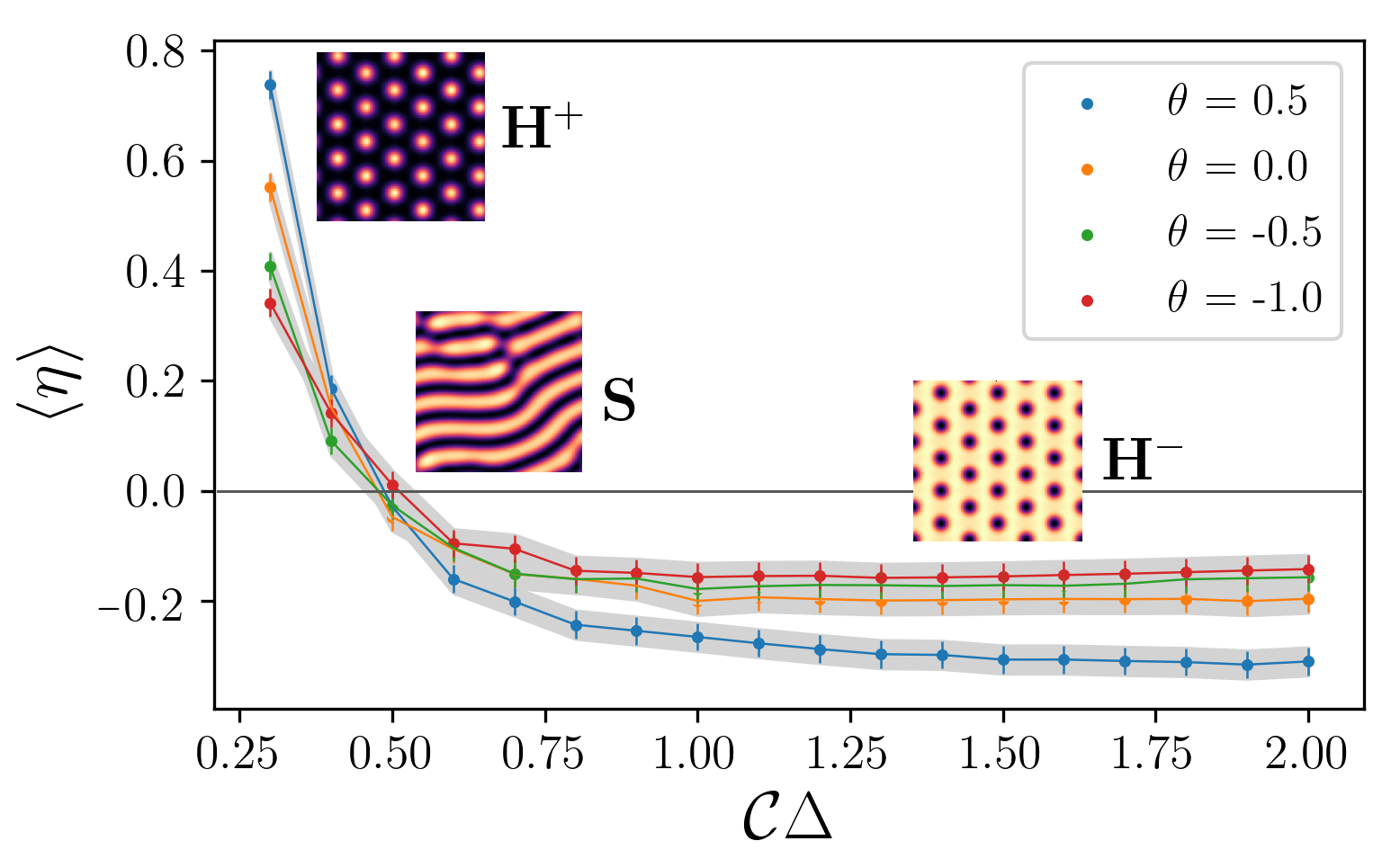}
    \caption{Average displacement $\langle \eta \rangle$ across the inversion symmetry point $\langle \eta \rangle = 0$ of the atom density $n_{\textrm{eq}}(\mathbf{r})$ from the homogeneous value $n_{\textrm{eq}} = 1$ for $\theta = -1, -0.5, -0.5$ at fixed $\sigma = 50$. Data are measured the steady state atom density according to Eq.~(\ref{fom}) by spanning values of $\mathcal{C}\Delta$ for a chosen value of $\theta$. The error bars reflect local variations of the value $\langle\eta\rangle$ across the domain $\Omega$.  }
    \label{invsim}
\end{figure}

As shown in Ref.~\cite{baio2021multiple} for a SFM configuration, the optomechanical instability displays structural transitions among patterned phases and a recovery of the inversion symmetry in dependence on the cloud susceptibility. The phenomenology of this mechanism is similar to cases where an external parameter is tuned such as, e.g., external fields \cite{krevsic2019inversion} or polarization balances \cite{Scroggie1996, aumann1997polarized}. For the present model, the linear susceptibility of the atomic cloud is encoded in the cavity cooperativity $\mathcal{C}$, introduced in Sec.~\ref{sec:level2}. Therefore, significant nonlinear behaviour beyond the effective-Kerr medium case (for fixed values of $b_0$) is expected when $\mathcal{C}\Delta$ lies below a certain value. Assuming $\mathcal{C}\Delta$ as a free parameter in our simulations, we span across the range $\mathcal{C}\Delta \in [0.25,2]$, corresponding to variations of $b_0 \in [10,100]$ for $\Delta \approx 100$ and $\tau = 0.2$ \footnote{We also assume a fixed $\sigma = 50$, providing $\Delta \approx 200$ for $T = 295\, \mu\textrm{K}$.}. Results are shown in Fig.~\ref{invsim} for different cavity detunings $\theta$, where the displacement of the steady-state atom density pattern $n_{\textrm{eq}}(\mathbf{r},\kappa t_{\textrm{max}})$ with respect to the homogeneous value $n_{\textrm{eq}} = 1$ is measured by the quantity:
\begin{align}
\langle\eta\rangle = \frac{1}{2}\left[\max_{\Omega} n_{\textrm{eq}}(\mathbf{r})+\min_{\Omega} n_{\textrm{eq}}(\mathbf{r})\right]  - 1.   
\label{fom}
\end{align}
As expected from the general notion of inversion symmetry, the figure of merit $\langle\eta\rangle$ used here is positive (negative) in correspondence of an $\mathbf{H}^{+(-)}$ atom density phase. Starting from the $\mathbf{H}^-$ phase already known from Refs.~\cite{Tesio2012, Tesio2013}, we observe a clear change in the symmetry of the self-organized light-atom pattern, roughly around a value of $\mathcal{C}\Delta = 0.5$. Moreover, stable $\mathbf{S}$ phases with possible defects are found in the vicinity of the point $\langle \eta \rangle = 0$. For $\mathcal{C}\Delta < 0.5$, one finds $\mathbf{H}^+$ states, extending the scenario from the SFM model in Ref.~\cite{baio2021multiple} to the ring cavity model. Interestingly, the critical $\mathcal{C}\Delta$ is independent from the cavity detuning $\theta$, further confirming the common origin of the transition, as stemming from the transport character of the nonlinearity. A rigorous characterization of the supercritical stability diagram in the space $(\mathcal{C}\Delta,\theta)$ by means of a weakly nonlinear analysis will be considered in a further study.

%\subsubsection{Dark light optomechanical cavity solitons}

\begin{figure}
    \centering
    \hspace*{-.25cm}\includegraphics[scale=.23]{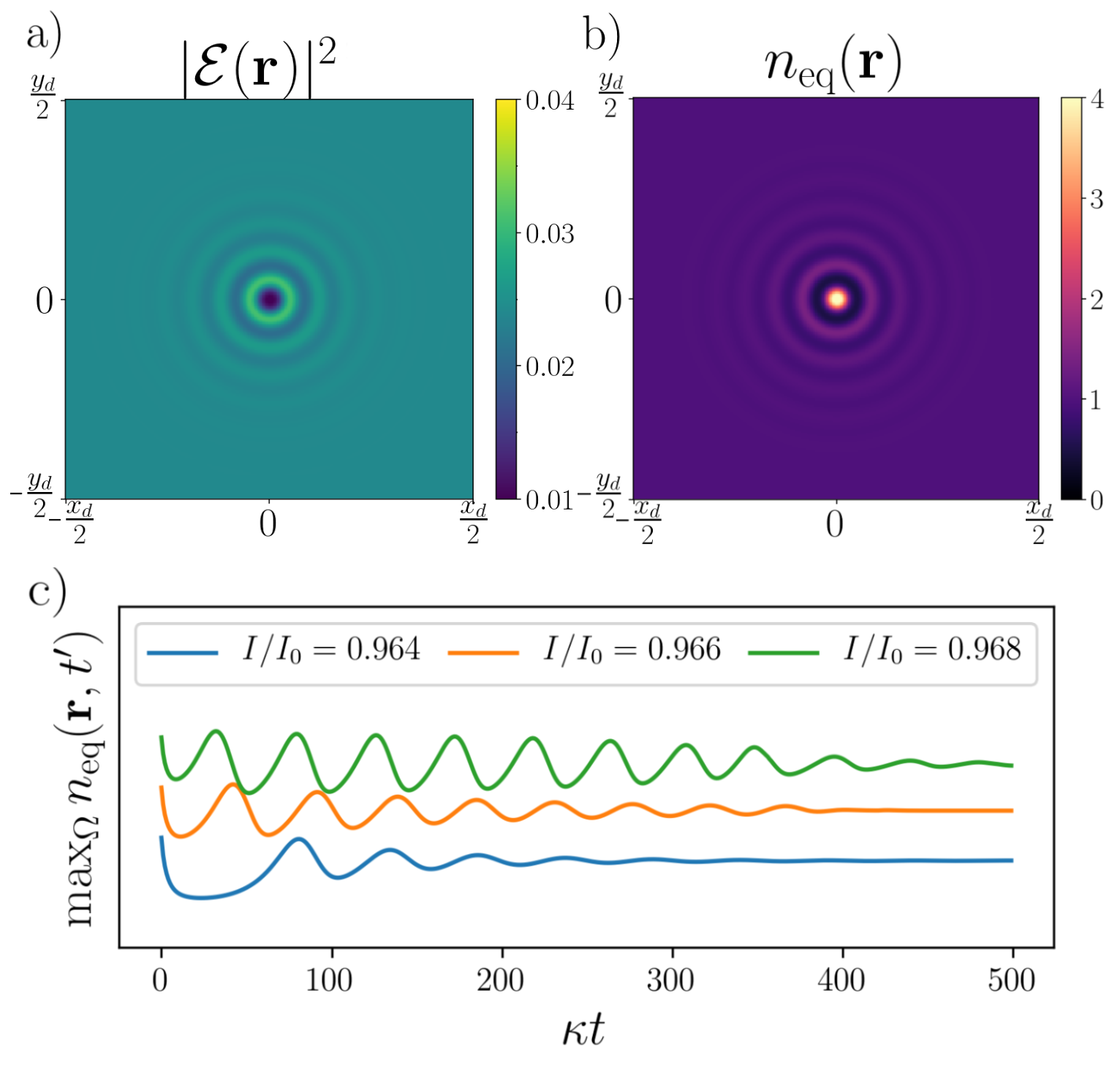}
    \caption{Dark light optomechanical CSs for blue light-atom detuning with plane-wave pumping. (a)-(b) Soliton spatial profiles with model parameters $I/I_0 = 0.968$, $\theta = -0.85$, $\mathcal{C}\Delta = 0.2$, $\sigma=100$. Oscillating soliton behavior obtained by tracking maximum density peak over time for pump values in the range  $I/I_0 \in [0.964,0.968]$. Each curve is vertically shifted for ease of understanding. The profiles in (a)-(b) are plotted in correspondence of an atom density peak maximum at $kt\approx 40$ for the corresponding green curve.}
    \label{oscil}
\end{figure}

The stability of $\mathbf{H}^+$ atom densities for blue detuning implies the existence of CSs characterized by atomic bunches self-trapped in a dark region of the optical field. This is shown in Fig.~\ref{oscil}~(a)-(b), with the characteristic diffraction rings \cite{baio2021multiple}. Such solitons are similar to the optical and matter-wave counterparts obtained with radially symmetric potentials \cite{kartashov2005stable, baizakov2006matter, kartashov2019stable}. Interestingly, we observe here regimes of weakly damped temporal oscillations, dependent on the input pump and shown in Fig.~\ref{oscil}~(c), by tracking the time evolution of the atom density peak $\mathrm{max}_{\Omega}\, n(\mathbf{r},t')$. The presence of such an oscillatory mode, excited by perturbations to the stationary soliton profile, reveals that optomechanical CSs may also display a Hopf instability in the vicinity of the current parameter regime \cite{skryabin1999interaction, firth2002dynamical}.

\begin{figure}
    \centering
    \hspace*{-.3cm}\includegraphics[scale=.20]{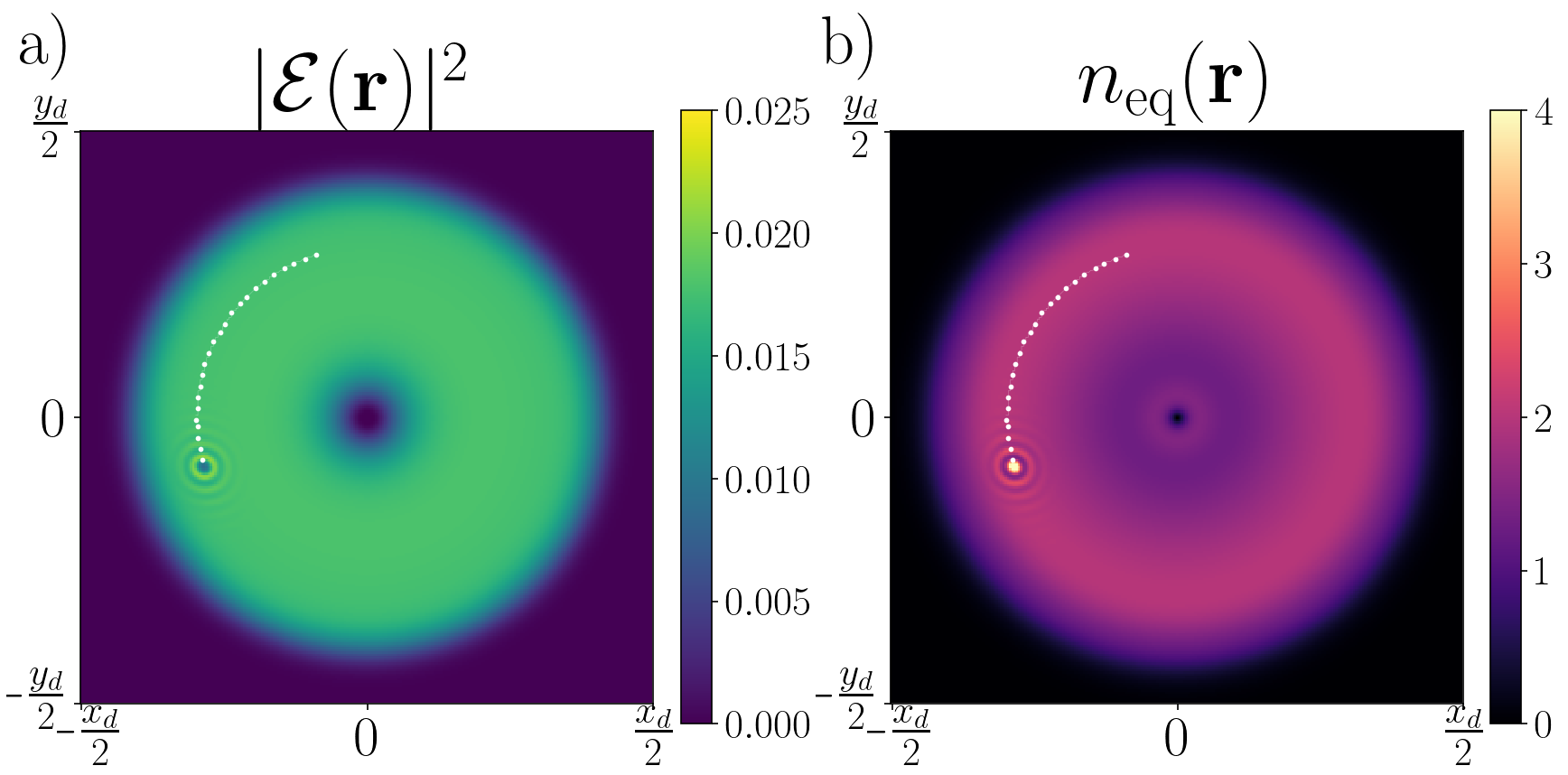}
    \caption{Counterclockwise rotation of a dark light optomechanical CS and its trajectory measured within a period of $\kappa t_{\mathrm{max}} = 10^3$ and higher OAM index $l=5$.  Parameters chosen as follows: $I/I_0 \approx 0.495$, $\theta = -0.43$,  $\mathcal{C} \Delta = 0.16$, $\sigma = 100$. An additional radial
    trap prevents atoms from accumulating in the dark regions of optical intensity.}
    \label{darksolrot2d}
\end{figure}

In the rest of this section, we address rotation of such solitons on a phase profile carrying OAM. First, a characteristic of the strong blue detuning regime  considered here is that, with a finite size pump, the atom density becomes negligible in the interaction region \footnote{This means that atoms simply tend to be pushed away from the input beam.}. Thus, we introduce an additional radial confinement in the atom density, achievable by further external trapping beams. Due to the higher-order rings visible from Fig.~\ref{oscil} (a)-(b), CSs in this regime interact strongly with the diffractive modulations of the homogeneous state below threshold. Such an effect is controlled by enlarging the domain size ($\approx35$ critical wavelengths) and smoothening the input field close to the singular point $\mathbf{r}=0$ \cite{yao2019control}. This is shown in Fig.~\ref{darksolrot2d}~(a)-(b), where the rotation velocity of the soliton is in once again in agreement with the predicted value in Eq.~(\ref{driftvel}). In Sec.~\ref{sec:level4}, we focus on interacting CSs and dynamical phenomena deriving from such interactions.

\section{\label{sec:level5} Multi-soliton interactions}

A general treatment of interacting CSs was derived in Refs. \cite{mcsloy2002computationally,vladimirov2002two} for the purely absorptive two-level case, based on neutral modes corresponding to translational invariance of the localized state \cite{obukhov1990self, maggipinto2000cavity}. Such an approach yields a hierarchy of equations of motion in gradient form, including the effect of many-body interactions \cite{scroggie2002self}. Those works predict stable bound states of two CSs at a set of preferred distances, mediated by their oscillatory tails. Those were experimentally observed for a sodium vapor in a SFM configuration \cite{schapers2000interaction}, where each pinning distance corresponds to an interaction potential minimum \cite{mcsloy2002computationally, vladimirov2002two}. Phase-structured input field can be used to induce soliton-soliton collisions, resulting into the formation of localized pattern spots or bound states of the light-atom CSs. This depends strongly on the input pump, which can inhibit the self-replication induced by the overlapping rings \cite{liehr2003replication}. In the rest of this section we show that OAM induced rotation can be used to probe multi-soliton interactions in a linear chain of CSs.
\begin{figure}
    \centering
    \hspace*{-.15cm}\includegraphics[scale=.21]{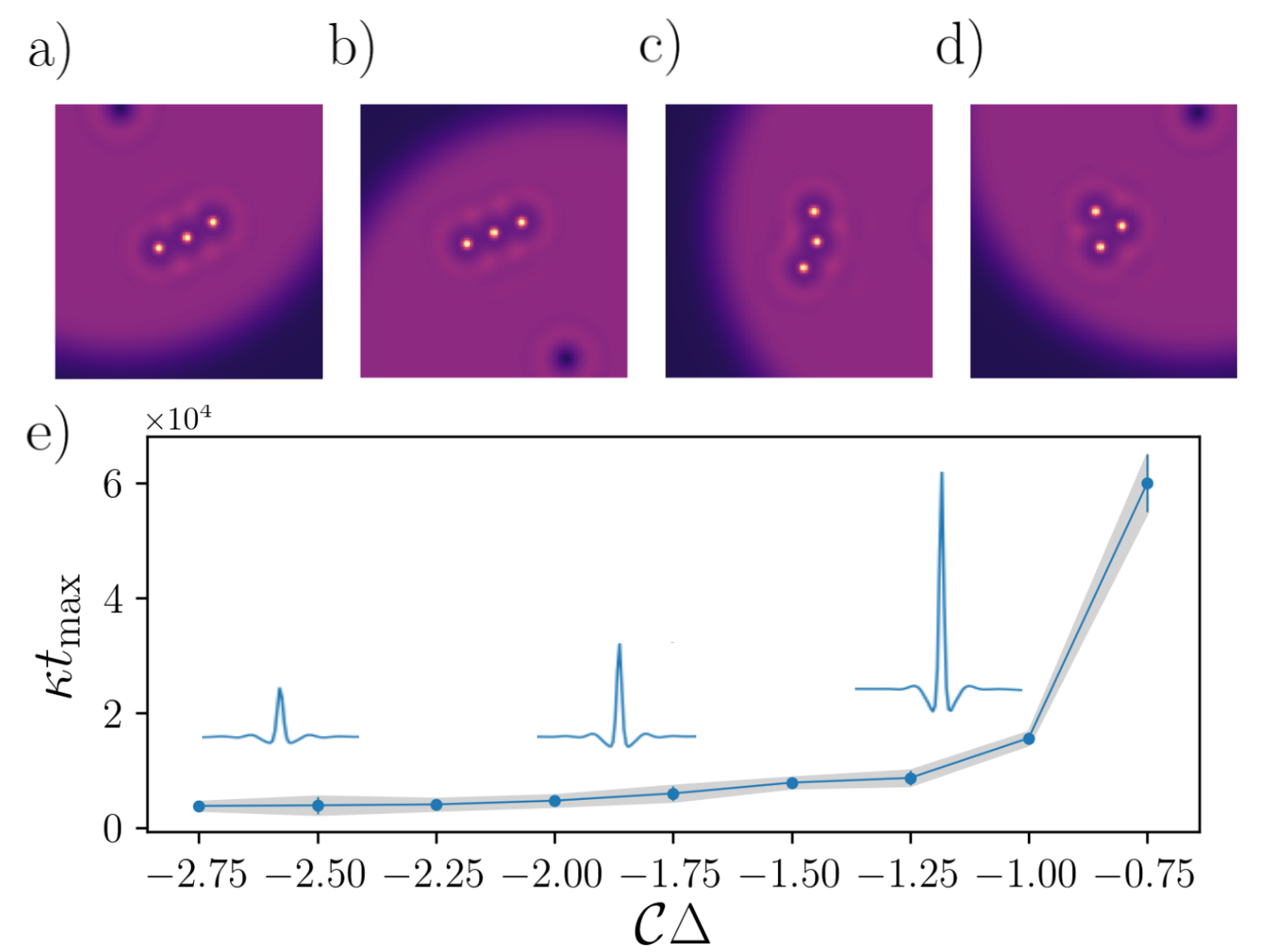}
    \caption{3-CS chain rotational dynamics with OAM index $l=1$. Rotating CSs initially excited in a bound state evolve into a triangle. Snapshots of atom density $n_{\mathrm{eq}}(\mathbf{r})$ at $kt=10^3$~(a), $kt=3\times 10^3$~(b), $kt=3.75\times 10^3$~(c), $kt=4.25\times 10^3$~(d) for $\mathcal{C}\Delta =-2.5$ and $\sigma = 25    $.~(e) Stability of the rotating 3-soliton chain together with corresponding single CSs profiles. This result suggests that the stability of CS chains increases with decreasing susceptibility $|\mathcal{C}\Delta|$. The error bars measure the observed duration of the chain collapse in units of $\kappa t$. }
    \label{chain}
\end{figure}

As discussed in Sec.~\ref{sec:level3}, in the presence of an phase gradient, a CS moves away from its initial position with a drift velocity dictated by the gradient itself. Thus, rotating or spiralling chains of two or more peaks can be constructed by periodically exciting travelling CSs with a purely azimuthal or combined radial + azimuthal phase such as in Eq.~(\ref{driftspir}). Chains of CSs orient themselves along the trajectories shown in Fig.~\ref{spir2}~(a)-(c). However, rotating chains are observed to be unstable in the long-term dynamics due to the local radial variation of the drift velocity and the system tendency to favor the $\mathbf{H}$-lattice \cite{vladimirov2002two}, as seen in Fig.~\ref{chain}~(a)-(d). The transient lifetime, during which linear CS chains are observed, is found to strongly depend on the susceptibility $\mathcal{C}\Delta$, meaning that soliton-soliton interactions also influence the stability of such states.  A detailed investigation of this aspect for a 3-CS chain in the rotating case, for different values of $\mathcal{C}\Delta$, is shown in Fig.~\ref{chain}~(e). We choose the 3-CS chain, where effects induced by the local phase gradient are enhanced. To evaluate the stability of the chain, we estimate the $\kappa t_{\textrm{max}}$ before it collapses into a triangle (See Fig.~\ref{chain}). Note that the triangular CS cluster continues to rotate at constant speed around the beam center. %The minimum observed transient is of the order $\kappa t \approx 4\times 10^3$ for $\mathcal{C}\Delta = -2.75$. 
The lifetime increases with $\mathcal{C}\Delta$ and it is practically infinite for $\mathcal{C}\Delta > -0.75$, meaning that the chains are stable.

The origin of such a behaviour can be traced back to the overlap between rings of interacting CSs. As visible from the insets in Fig.~\ref{chain} (e), for $\mathcal{C}\Delta > -1$, the peaks corresponding to higher-order rings of a single CS are smaller with respect to the central peak. Therefore, one expects CSs interactions to lose relevance in the overall dynamics, partially explaining the results in Fig.~\ref{chain}. This can be investigated quantitatively by means of effective Hamiltonian approaches~\cite{parra2017interaction}. Those could unveil the presence of configurational minima besides the triangular cell.

\section{\label{sec:level5} Concluding remarks}

We demonstrated controllable motional states of transverse optomechanical localized structures in a longitudinally pumped ring cavity, where the nonlinear medium is given by a cloud of laser-cooled atoms. This motion arises from phase structured input fields carrying OAM, generating atomic transport via an optomechanical instability \cite{Baio2020}. In particular, by means of numerical studies, we addressed complex rotational and spiralling trajectories of optomechanical CSs by tuning the radial and azimuthal dependencies of the input profile. We also reported structural transitions among patterned phases with different symmetry and rotational motion of corresponding dark-light CSs. Finally, we explored CSs interactions in rotating 3-CS linear chains, providing evidence that they play a crucial role in the stability of such bound states. 

A direct extension of interest for the present work is the study of two or more optomechanical CS collisions beyond the overdamped regime, unveiling potential effects of the transient dynamics in the momentum distribution or connections to supersolid droplets in the ultracold limit \cite{PhysRevE.104.044201}. Similar transverse localized states are also found in the optomechanical coupling of discrete arrays of oscillating mirrors \cite{ruiz2020spatial, PhysRevA.93.033850}.

Finally, our results suggest the possibility of transporting self-trapped atoms by means of CS motion in arbitrary time-dependent phase profiles \cite{pedaci2006positioning, cleff2008gradient}. All such studies are of potential relevance for the realization of novel atomtronic devices \cite{amico2021atomtronic, amico2021roadmap}.

\begin{acknowledgments}
All authors acknowledge financial
support from the European Training Network ColOpt, which
is funded by the European Union (EU) Horizon 2020 program
under the Marie Skłodowska-Curie Action, Grant Agreement
No. 721465.
\end{acknowledgments}

\appendix

% \section{Appendixes}

% To start the appendixes

%\nocite{*}

\bibliography{short, references}% Produces the bibliography via BibTeX.

\end{document}